\documentclass{article}
\usepackage{amsmath,amssymb}
\usepackage{amsfonts}
\usepackage{graphicx}
\usepackage{setspace}
\usepackage{color}

\newcommand {\be}{\begin{eqnarray}}
\newcommand{\ee}{\end{eqnarray}}

\begin{document}

\begin{flushright} 
LAPTH-008/17 \\ % \\ KEK-Cosmo-, 
KEK-TH-1969
\end{flushright}
\vspace*{5mm}

\begin{center}
{\bf
\begin{spacing}{1.0}
\LARGE QCD-Electroweak First-Order Phase Transition in a Supercooled Universe
\end{spacing}
}
\vspace*{10mm}
{\large
Satoshi Iso,${}^{\; a}$ Pasquale D. Serpico,${}^{\; b}$ and Kengo Shimada${}^{\; b}$
}
\vspace{10mm}

{\it \small
$^a$ Institute of Particle and Nuclear Studies, High Energy Accelerator Research Organization (KEK) and Graduate University for Advanced Studies (SOKENDAI), Oho 1-1, Tsukuba, Ibaraki 305-0801, Japan \\
$^b$  LAPTh, Universit\'e de Savoie Mont Blanc and CNRS, 74941 Annecy Cedex, France
}

\end{center}

\vspace{10mm}

\begin{abstract}
If the electroweak sector of the standard model is described by classically conformal dynamics, the early Universe evolution can be substantially altered.
It is already known that---contrarily to the standard model case---a first-order electroweak phase transition may occur. Here we show that, depending on the model
parameters, a dramatically different scenario may happen: A first-order, six massless quark QCD phase transition occurs first, which then triggers the electroweak symmetry breaking.
We derive the necessary conditions for this dynamics to occur, using the specific example of the classically conformal $B-L$ model.  In particular, relatively light weakly coupled
particles are predicted, with implications for collider searches. This scenario is also potentially rich in cosmological consequences, such as renewed possibilities for electroweak baryogenesis, altered dark matter production, and gravitational wave production, as we briefly comment upon.
\end{abstract}

\newpage 
%%%%%%%%%%%%%%%%%%

\section*{Introduction}
Despite the recent discovery of the Higgs boson $h$, we still have little clue on the 
physics beyond the standard model (SM) at or above the electroweak (EW) scale. In the past decades,
model building has been mostly focusing on supersymmetric or Higgs compositeness scenarios at (sub-)TeV scale,
motivated by the naturalness of the Higgs mass value. These approaches are, 
however, under strain, due to tighter and tighter experimental bounds on the masses of new particles, notably of colored ones, predicted
in such models.
Hence there is renewed motivation to explore alternatives, notably theories including very weakly coupled particles, possibly lighter than SM ones.

An old  theoretically appealing idea is that EW symmetry breaking (EWSB) is induced by radiative corrections to the Higgs potential, manifesting conformal symmetry at tree level if its mass term vanishes~\cite{Coleman:1973jx}.
This possibility is nowadays excluded in the SM due to the measured values of its parameters,
but it might be viable in classically conformal (CC) extensions of the SM where at least an additional scalar field $\phi$ is introduced, a requirement anyway needed to account for neutrino masses~\cite{Meissner:2006zh} at least if originating through a seesaw mechanism.
In a frequently considered implementation of this scenario~\cite{Foot:2007as}--\cite{Holthausen:2009uc}, 
it has been noted that  the phase transition (PT) breaking the EW symmetry tends to be strongly first-order.
Then a significant supercooling below the critical temperature 
and a relatively long time scale for bubble percolation 
are implied, and thus a sizable gravitational wave (GW) 
production and possibly electroweak baryogenesis (EWBG)~\cite{Espinosa:2008kw}--\cite{Sannino:2015wka} are expected.
See also \cite{Hur:2011sv}--\cite{Konstandin:2011dr} for other realizations with hidden strong dynamics.

Despite their conceptual simplicity, CC models may lead to an even more fascinating possibility: If the supercooling is maintained down to temperatures lower than the QCD critical temperature $T^{\rm QCD}_c$, chiral symmetry breaking ($\chi$SB) occurs spontaneously via quark condensation, $\langle \bar{q} q \rangle \neq 0$.
Contrarily to the current phase of the Universe, all the quarks were then massless, as initially the scalar fields have no vacuum expectation value (VEV). 
The chiral symmetry is thus broken from $SU(6)_L \times SU(6)_R$ to $SU(6)$, and the associated PT is then first-order~\cite{Pisarski:1983ms}.
At the same time, $\langle \bar{q} q \rangle \neq 0$ also breaks the EW symmetry, since $\bar{q}q$ is an $SU(2)_L$ doublet with a nonvanishing $U(1)_Y$ charge,
a situation that has been recently  considered as relevant only in ``gedanken worlds''~\cite{Quigg:2009xr}.  
Furthermore, when $\chi$SB occurs, the Yukawa couplings $y_i$ with the SM Higgs $h$ generate a linear term $y_i h \langle \bar{q}_i q_i \rangle /\sqrt{2}$,
tilting the scalar potential along the $h$ direction.
This tilt destabilizes the false vacuum at the origin
and the Higgs acquires a VEV {\it at the QCD scale}.
A similar possibility within the SM had been already entertained by Witten~\cite{Witten:1980ez}
but is long since excluded.
A couple of decades ago, it was occasionally reconsidered in SM extensions with a dilaton field~\cite{Buchmuller:1990ds} or in applications to EWBG~\cite{Kuzmin:1992up}.
The goal of this Letter is to show that it currently remains a concrete possibility 
in CC extensions of the SM,  implying a qualitatively different history of the early Universe.  In the following, we focus on characterizing the conditions for a QCD-induced EWSB, commenting upon some  particle physics and cosmological consequences of such a scenario. \\

\section*{The model}
For definiteness, 
let us consider the CC $B-L$ extension of the SM~\cite{Iso:2009ss}, where the $B-L$ symmetry is gauged, with gauge coupling $g$.
Besides the SM particles, the model contains a gauge boson $Z'$,  a  scalar $\Phi$ with $U(1)_{B-L}$ charge 2, 
and three right-handed neutrinos (RH$\nu$) canceling the $[U(1)_{ B-L}]^3$ and gravitational anomalies.
The CC assumption requires that the scalar potential $V$, within renormalizable field theories,
  involving $\Phi$ and the Higgs doublet  $H$ has no quadratic terms and is given,  up to a constant term, by $ \hspace*{0mm} V(H,\Phi) =
 \lambda_h |H|^4 + \lambda_{\rm mix} |H|^2 |\Phi|^2 + \lambda_\phi |\Phi|^4 .$ 
It is then assumed that  the $B-L$ symmetry is radiatively broken by the Coleman-Weinberg (CW) mechanism~\cite{Coleman:1973jx}, which triggers the EWSB. 
For this, the scalar mixing $\lambda_{\rm mix}$
is required to take a small negative value.
Here we summarize approximate key formulas \cite{Suppl-1}.
At one loop, the CW potential along the potential valley is approximated by
$ V_{\rm CW}^{\rm v} (\phi)= V_0 + B \phi^4 [ \ln (|\phi|/M) -1/4 ]/4$  where $\phi  \approx \sqrt{2}|\Phi|$. 
If the condition $B>0$ is satisfied, the effective potential has a global minimum at $\phi=M$, the scale generated radiatively via the RG evolution of the quartic coupling (see \cite{Iso:2009ss}).
Note that the zero temperature one-loop CW potential $V_{\rm CW}^{\rm v}$ has no  barrier between
$\phi=0$ and $M$.
At the global minimum, various particles acquire masses: 
$m_{Z'}=2 g M$, $m_{\phi} = \sqrt{B} M$, and
  $m_{N_i}=Y_i M/\sqrt{2}$ for the RH$\nu$'s whose Yukawa couplings are $Y_i$.
The constant term  $V_0 = BM^4 /16$ is chosen so that $V_{\rm CW}^{\rm v}(M)=0$.
The coefficient $B$ is approximately given by 
$B \approx {\sf c}_0^4 [ 3 (2 g)^4 -D\lambda_{\rm mix}^2 -2 \  {\rm Tr}(Y/\sqrt{2})^4 ]/ 8 \pi^2$ 
where Tr is the trace over three RH$\nu$ flavors, ${\sf c}_0^4 \approx (1+\lambda_{\rm mix}/\lambda_h)$, and $D\simeq 41$ (see Supplemental Material).
The coefficient ${\sf c}_0^4$ and  $D\neq -1$ represents the admixture of $h$ and $\phi$ along the valley.
Because of $\lambda_{\rm mix}<0$,  the Higgs field $h=\sqrt{2} |H|$ has a  minimum at $v=(|\lambda_{\rm mix}|/2 \lambda_h)^{1/2} M$, 
 identified with  the Higgs VEV $246\, {\rm GeV}$. 
If $M \gg v$, the Higgs mass  is given by $m_{h} \approx \sqrt{|\lambda_{\rm mix}|}\, M$.
In spite of the large hierarchy $M \gg v$,  the scalar $\phi$ is generically light, 
$m_\phi\lesssim 10 \, m_h$  \cite{Iso:2012jn}. 
In the following, for simplicity we shall require $g\gtrsim 10^{-4} (m_{Z'}/{\rm TeV})^{1/4}$ so that $\phi$ and $Z'$ are thermalized before the supercooling stage,
although some of our conclusions may be true in a broader parameter space.
 
\section*{Hypercooling in the EW sector}
A CC system has peculiar thermodynamic properties.
To see this, let us focus on a model accounting only for the fields
$\phi$ and  $Z'$ and at the leading order in the high-temperature expansion (See the Supplemental Material for some considerations on the quality of these approximations).
The effective potential 
is thus approximated as (see, e.g., Ref. \cite{Dolan:1973qd})
$ V(\phi)=(c_2/2)T^2 \phi^2 - (c_3/3) T \phi^3 +(B_{Z'}/4) \phi^4 \ln (T/\hat{\mu})$
where  $\phi$-independent terms are dropped.
The coefficients are given by 
 $c_2=g^2$, $c_3=   6g^3 /\pi$, $B_{Z'} = 6g^4 /\pi^2$, and $\hat{\mu} =m_{Z'}\,e^{\gamma_E-1/2}/4\pi$.
At sufficiently high $T$, the quadratic term dominates and the only minimum
of the potential  is at $\phi =0$: $B-L$ symmetry is restored.
 We study the cosmological evolution of the Universe with an initial condition $\phi=0$,
which is naturally realized after the large field inflation as discussed, e.g., in Refs. \cite{Linde:1981mu}--\cite{Iso:2015wsf}.
When $T$ drops below the critical temperature $T_c \sim m_{Z'}$,
defined by the condition  ${\cal C}(T_c) \equiv  9 c_2 B \ln (T_c/\hat{\mu}) / (2 c_3^2) = 1$,
the nontrivial minimum of the potential at $\phi_c = 3 T c_2 /c_3 \lesssim M$ has a lower energy compared to the false vacuum $\phi =0$.
However, due to the CC assumption,  
the coefficient of the quadratic term $c_2$ is always positive and
the false vacuum remains the local minimum even at $T \ll T_c$: 
the thermal potential barrier never disappears.
Hence, the Universe with the Hubble expansion rate 
$ {\cal H}=\sqrt{V_0/3\,m_{\rm pl}^2}$ is supercooled down to a very low temperature, where $m_{\rm pl} \approx 2.4 \times 10^{18}$ GeV is the reduced Planck mass.

%%%%%%%%%%%%%%%%%%%%%%%%%%%%%%%%%%%%%%%%
\begin{figure}[t]
\centering
\includegraphics[width=1 \linewidth, bb=0 0 830 334]{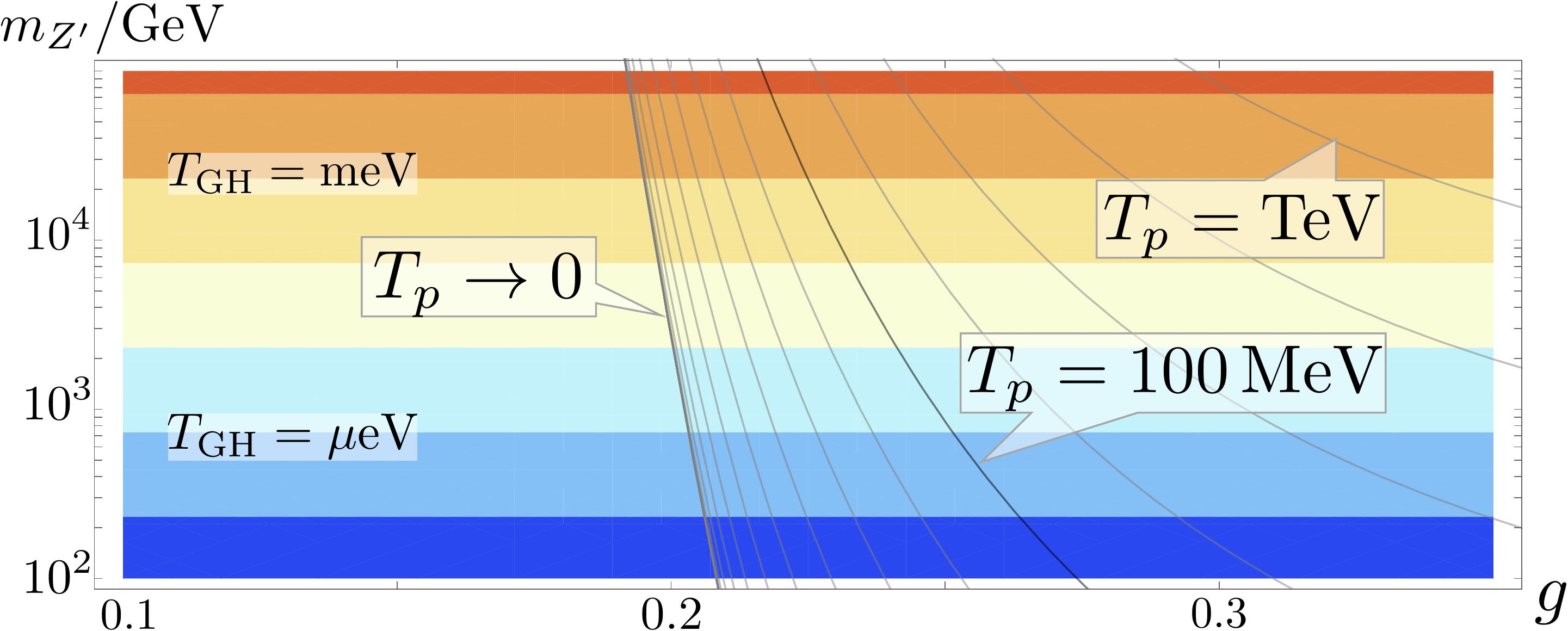}
\caption{\label{Fig:Tp}
Contour plot of the percolation temperature $T_p$ (black lines) as a function of $g$ and $m_{Z'}$. 
The horizontal color bands show the temperature
 $T_{\rm GH} \equiv {\cal H}/2\pi$.
}
\end{figure}
%%%%%%%%%%%%%%%%%%%%%%%%%%%%%%%%%%%%%%%%%%%

The Universe may eventually percolate into the true vacuum via bubbles nucleated by quantum tunneling. 
The percolation temperature $T_p$ can be estimated by using the tunneling rate
$\Gamma \approx T^4 e^{-S_3/T}$. 
In the present model, the critical bubble's action is given by
$S_3/T\approx A [1-2\pi {\cal C}(T)/9]^{-1}$,
where $A= 43.7 \, c_2^{3/2}/c_3^{2} \propto g^{-3}$ and
 ${\cal C}(T) =(3/4)  \ln(T/\hat{\mu})$ for $T<\hat{\mu}$, consistently with results in Ref. \cite{Dine:1992wr} for a non-negative quartic coupling. 
Thus, for $g \ll 1$ the tunneling rate becomes very small.
The fraction of space remaining in the false vacuum at a given temperature $T<T_c$ is given by $p(T)=e^{-I(T)}$, where $I(T)$ is defined by the probability that a single bubble of true vacuum is nucleated in the past (see Ref. \cite{Guth:1981uk}). 
The percolation temperature  $T_p$ is then defined by the condition $I(T_p) = 1.$
In Fig.~\ref{Fig:Tp}, we plot $T_p$  as a function of $g$ and $m_{Z'}$. 
Because of the (weakly $T$-dependent) behavior $S_3/T \propto g^{-3}$,  percolation does not occur for $g \lesssim 0.2$. 
Eventually, the transition to the true vacuum would occur when the de Sitter fluctuation ${\cal O}(1) \times T_{\rm GH}$ becomes comparable to the width of the barrier, $\Delta \phi\sim T/g $, where $T_{\rm GH}\equiv{\cal H}/2\pi$ is the Gibbons-Hawking temperature.
This does not happen until temperature becomes very low when de-Sitter fluctuation destabilizes the false vacuum,
a condition that we dub {\it hypercooling}.
Note that the width of the barrier is evaluated as $\Delta \phi\sim T/g$ and implies that $m_{Z', {\rm eff}}< g\Delta \phi\sim T$, where the high-$T$ expansion is barely justified.
Our conclusion remains qualitatively correct in a more realistic treatment,  e.g., using the full thermal potential without high-$T$ expansion \cite{Suppl-2}.\\

\section*{QCD-induced EWSB}
If the percolation temperature of the $B-L$ sector 
is lower than the QCD critical temperature
 $T_c^{\rm QCD}$ and if the de Sitter fluctuation $\sim T_{\rm GH}$ is negligible compared to the QCD scale, the previous model cannot be trusted anymore to describe the dynamics,
since the CC condition is actually broken by QCD via dimensional transmutation, i.e., confinement and $\chi$SB. 
At the false vacuum, all the quarks are massless.
{\it QCD with $N_f=6$ massless quarks (or five massless and one massive near the false vacuum) has a first-order PT}~\cite{Pisarski:1983ms}, with $T_c^{\rm QCD}$ somewhat lower than that in the SM, e.g., 85 MeV in Ref. \cite{Braun:2006jd}.
Contrarily to the previously discussed case, the QCD PT 
 is expected to occur at  $T_n^{\rm QCD}$ only mildly below $T_c^{\rm QCD}$, 
 because QCD has a dynamical scale $\Lambda_{\rm QCD}$.
 We can check that hypercooling does not take place, e.g., by using the Polyakov-quark-meson model~\cite{Herbst:2013ufa}.

When the QCD PT occurs, namely, when the chiral condensates form, 
a linear term $\sum_i y_i \langle \bar{q}_i q_i \rangle h /\sqrt{2}$
is generated in the Higgs potential, and a new local minimum $h=v_{\rm QCD}\sim  {\cal O}(100)$ MeV emerges. 
At this minimum, quarks (even the top quark) acquire very light masses $m_{q_i}=y_i v_{\rm QCD}/\sqrt{2} \lesssim \Lambda_{\rm QCD}$.
Thus, all the $N_f=6$ quarks are expected to form a chiral condensate 
$\langle \bar{q}_i q_i \rangle$. 
The top Yukawa coupling $y_t$ sets the size of the linear term in the Higgs potential, i.e., the local minimum of the Higgs potential is estimated as 
$v_{\rm QCD} =(y_t  \langle \bar{t} t \rangle/\sqrt{2}\lambda_h )^{1/3}$. 
Note that the top behaves similarly to the strange quark in the present Universe, which has a mass $m_s \sim 100$ MeV comparable with 
the QCD scale, but whose condensate is of the same order as the up (or down) quark one. 

Also, the $SU(2)_L \times U(1)_Y$ gauge symmetries are spontaneously broken, and linear combinations  of the pions and the ordinary Nambu-Goldstone components of the Higgs field 
are eaten by the massive gauge bosons. Thus, {\it EWSB is triggered by the first-order QCD PT}.
\\

%%%%%%%%%%%%%%%%%%%%%%%%%%%%%%%%%%%%%%%%%%%%%%
\begin{figure}[t]
\centering
\vspace{-0.4cm}
\includegraphics[width=1 \linewidth, bb=0 0 757 335]{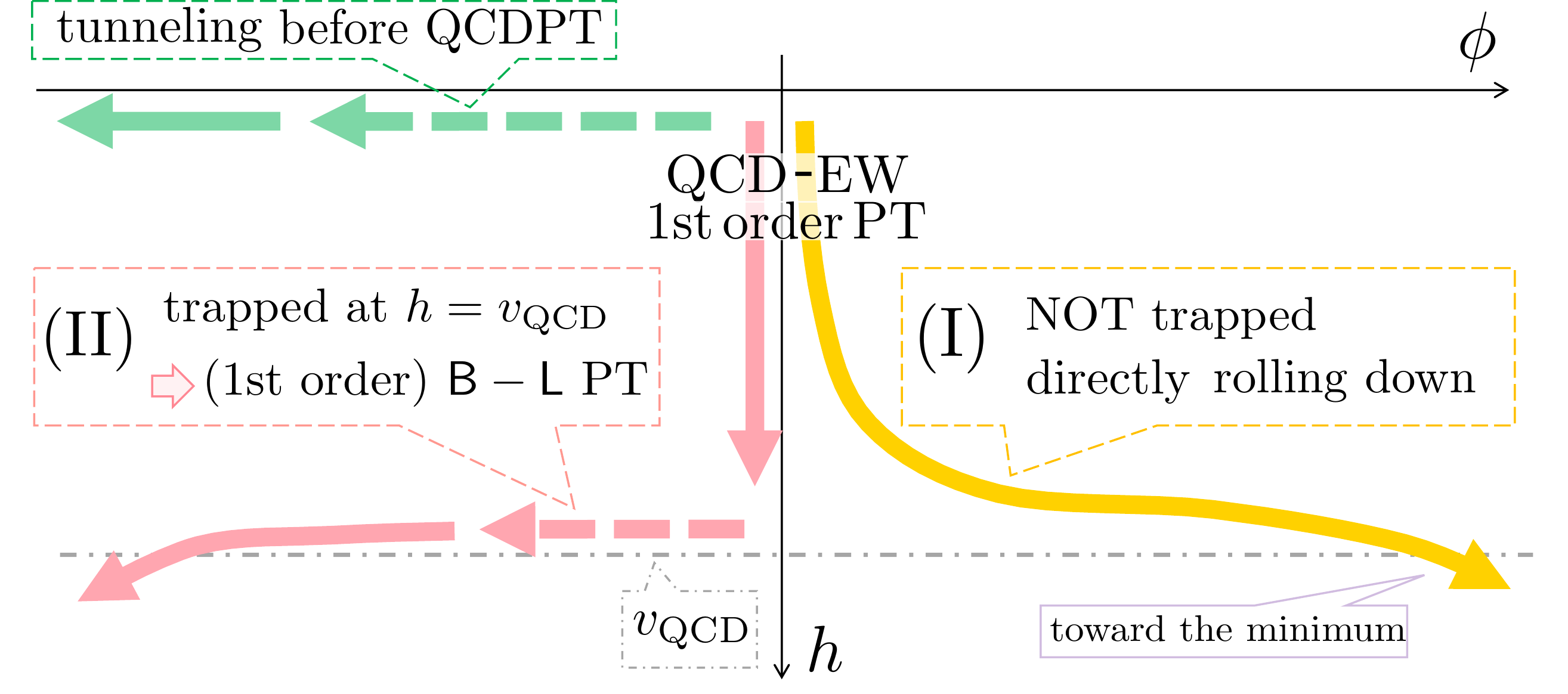}
\caption{\label{Fig:phases} 
Possible trajectories of the  scalar fields $(\phi, h)$ in the early Universe. All start from the origin $(0,0)$.}
\end{figure}
%%%%%%%%%%%%%%%%%%%%%%%%%%%%%%%%%%%%%%%%%%%%%%
%%%%%%%%%%%%%%%%%%%%%%%%%%%%%%%%%%%%%%%%%%%%%%%
\begin{figure}[t]
\centering
\vspace{-0cm}
\includegraphics[width=1.0 \linewidth, bb=0 0 882 369]{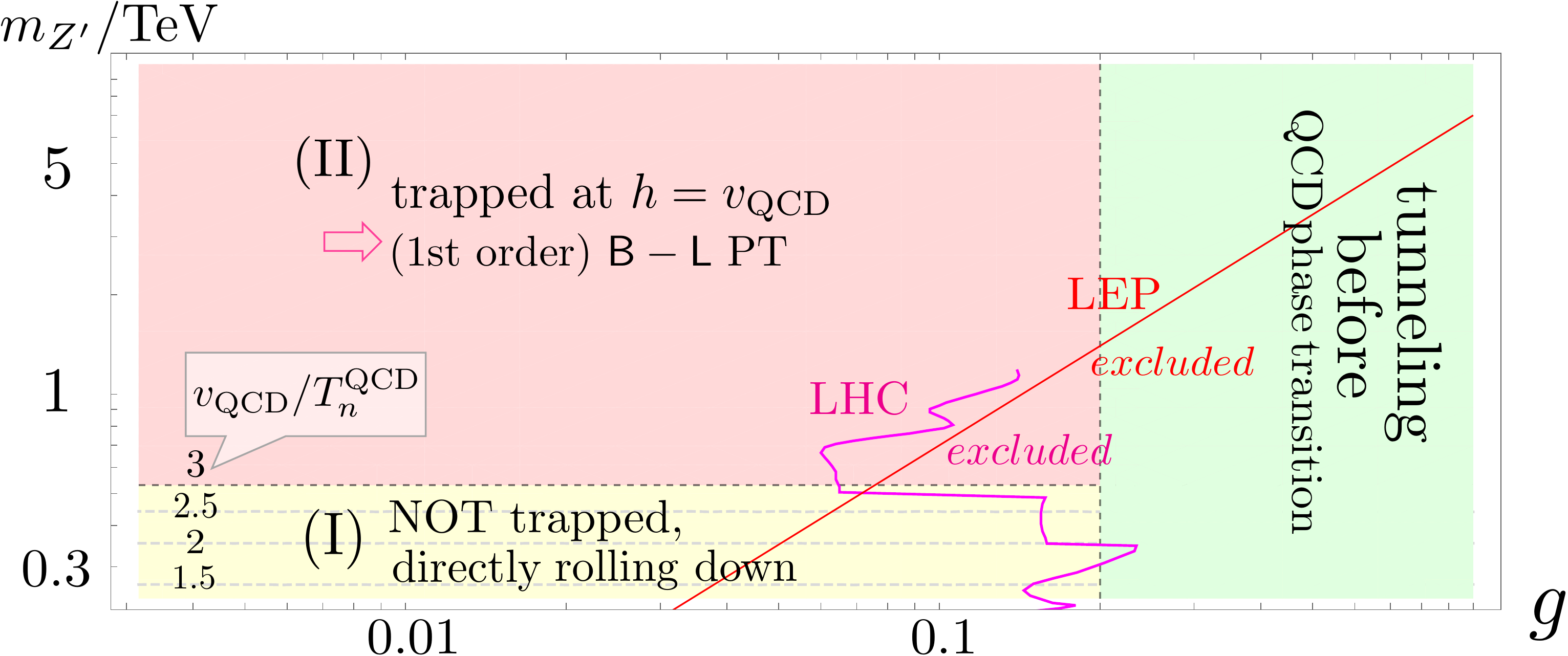}
\caption{\label{Fig:parameters}  Schematic cosmological histories for the parameter
region $m_{Z'}\sim {\rm TeV}$, assuming $m_N=0$ and $v_{\rm QCD}/T_{n}^{\rm QCD} = 3$.
The horizontal thin dashed lines are the boundaries between (I) and (II) for $v_{\rm QCD}/T_{n}^{\rm QCD} = 1.5,\, 2,\, 2.5$.
Below the lower edge, $B>0$ is violated.
For reference, we  plot the LEP bound \cite{Carena:2004xs} (red line) and some LHC bounds from Fig.6 in~\cite{CMS:2017dhi} (magenta line).}
\end{figure}
%%%%%%%%%%%%%%%%%%%%%%%%%%%%%%%%%%%%%%%%%%%%%%

\section*{Histories of the early Universe}
Different histories of the early Universe, i.e., different trajectories of the scalar fields, are possible as
in Fig.~\ref{Fig:phases}, depending on different values of the parameters $(g, m_{Z'})$ as in Fig.~\ref{Fig:parameters}.
If the percolation temperature $T_p$ is higher than the QCD scale $\Lambda_{\rm QCD} \sim 100$ MeV, 
$\phi$ field tunnels  into the true vacuum before the
QCD PT (green line in Fig.~\ref{Fig:phases}). 
A strong first-order PT takes place, and a sizable production of gravitational waves is expected \cite{Jinno:2016knw}. 
From Fig.~\ref{Fig:Tp}, such a possibility is realized for sufficiently strong gauge coupling $g \gtrsim 0.2$ as in  the green region in Fig.~\ref{Fig:parameters}.

If  $g \lesssim 0.2$, the QCD-induced EWSB occurs after de Sitter expansion 
with an $e$-folding $\sim \ln(T_i/T_n^{\rm QCD})$.
$T_i \equiv (30\,V_0/  \xi \pi^2)^{1/4}$ is the temperature when  the  expansion starts,
 and  $\xi \gtrsim 110$ is number of degrees of freedom in the extended SM. 
 After the QCD-induced EWSB,  if the quadratic term 
$
(c_2 T^2 +\lambda_{\rm mix} h^2/2)\phi^2 /2
$
at $T=T_n^{\rm QCD}$ and  $(\phi,h)=(0, v_{\rm QCD})$  is positive, the field $\phi$ is trapped [dubbed scenario (II)].
 If it is negative, $\phi$ rolls down to the true minimum $(\phi,h)=(M,v)$,  which we name scenario (I).
The trajectories are drawn  in Fig.~\ref{Fig:phases}. 
The trapping condition translates into $6 m_{Z'}^2 +  \text{Tr} (m_N^2) > 12 m_h^2 \  (v_{\rm QCD}/T_n^{\rm QCD})^2 $,
and is reported in pink in Fig.~\ref{Fig:parameters}.
Then thermal inflation occurs: As $T$ drops, the field tunnels and starts rolling around $T=\sqrt{ |\lambda_{\rm mix}|/(2c_2)} v_{\rm QCD}$ when the coefficient of the quadratic term vanishes.
If, instead, the trapping condition is violated,
$\phi$ freely rolls down \cite{trapping} [scenario (I),  yellow region in Fig.~\ref{Fig:parameters}].
On top of it, the fate of the Universe is also controlled by the slow roll condition  $|\eta | = m_{\rm pl}^2 |V''|/V_0 <1$ at $(0,v_{\rm QCD})$.
Namely, if $g \lesssim 10^{-2} \, (m_{Z'}/{\rm PeV})^3$ is satisfied, an inflationary expansion takes place after the phase transition.

Since the CW mechanism requires $B>0$,  i.e.  $3 m_{Z'}^4 > 2 \text{Tr}(m_N^4) + D m_{h}^4$, the necessary condition for scenario (I) reads $v_{\rm QCD}/T_n^{\rm QCD} > [D/12]^{1/4}\simeq 1.36$, i.e. $\langle \bar{t} t \rangle^{1/3} \gtrsim 0.77\, T_n^{\rm QCD}$. 
In the standard  two-flavor QCD case,   $\langle \bar{q} q \rangle^{1/3}/T_n^{\rm QCD}\simeq  0.5$, and one falls a little short of this condition. However, pending dedicated lattice studies, in our framework we cannot exclude that this inequality is actually satisfied. Anyway, as long as one is near the condition $3 m_{Z'}^4 \gtrsim D m_{h}^4$, either the scenario (I) is realized or a scenario (II) with very shallow trapping, i.e., very short inflation and a fast transition to the true vacuum.
This parameter space provides ideal conditions to observe, e.g., relics from the QCD-induced EWSB, limiting dilution.
It is also the regime where the $B-L$ gauge boson is predicted at the EW scale and RH$\nu$'s and the $B-L$ scalar below it,
which makes the model amenable to direct collider probes like the LHC  and, up to $\sim 4\,$GeV mass for RH$\nu$, also SHiP: See, for instance, Ref. \cite{Batell:2016zod} for a forecast study, reporting a sensitivity down to $g\sim {\cal O}(10^{-3})$.\\

\section*{Cosmological consequences}
We list a few cosmological consequences of the above scenario.

\begin{figure}[t]
\centering
\includegraphics[width=1 \linewidth, bb=0 0 874 415]{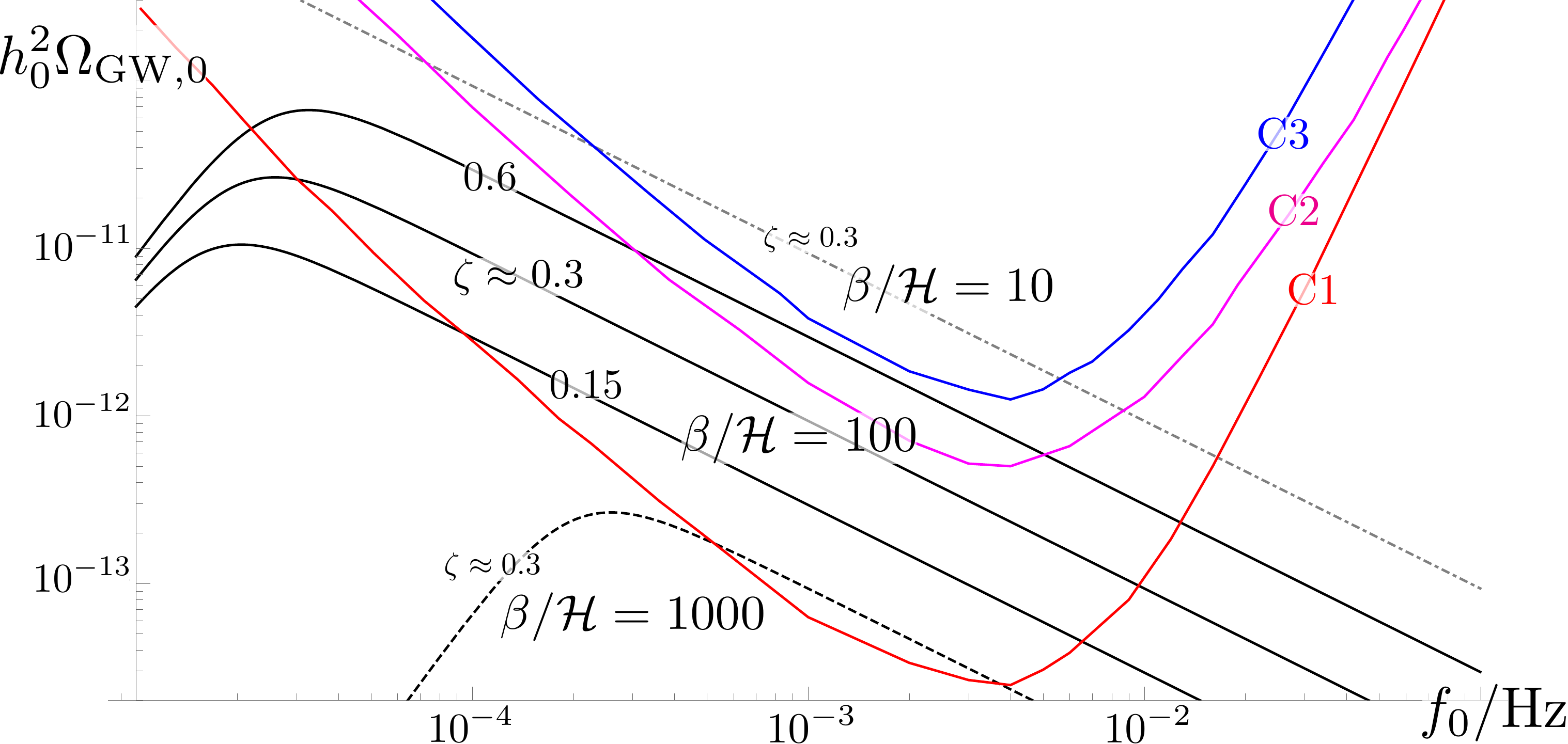}
\caption{\label{Fig:GW}The GW power spectrum for $\beta/{\cal H}=10,100,1000$.
We chose $T_i=10\,  {\rm GeV}$ and $\zeta=0.6,\, 0.3,\, 0.15$. 
The sensitivity of three configurations foreseen 
for the space mission LISA~\cite{Caprini:2015zlo} are also shown.}
\end{figure}

(i) The temperature after the PT is limited to $T_i \lesssim 20 \, {\rm GeV}$ in scenario (I).
Hence,  particles with mass $m \gtrsim {\cal O} (10) \times T_i$ (such as many dark matter candidates) cannot be thermally produced.
The viability of different types of dark matter candidates obtained via alternative production mechanisms should be thus revisited (see, e.g., Ref. \cite{Konstandin:2011dr}). 

(ii) {\it Cold} EWBG might take place, which has been argued to be a generic opportunity offered by  the supercooling stage ending with the first-order PT~\cite{Konstandin:2011ds}. An interesting possibility is a QCD axion extension \cite{Servant:2014bla}. As discussed in the {\it standard} EWBG context~\cite{Kuzmin:1992up}, 
the EWPT triggered by the $\chi$SB ``optimizes'' the efficiency of the strong $CP$ violation to that purpose. Of course, our scenario is very specific, and a modification of the EWSB dynamics has profound implications on several ingredients of the EWBG scenario, like the sphaleron energy and the necessary $CP$ violation.

(iii) Note that 
the $e$-folding $\ln (T_i/T_n^{\rm QCD})$ gained during the late inflationary period is small and unrelated to the one probed via cosmic microwave background (CMB) fluctuations.
This is a welcome consequence of our model, since  small-field inflations with simple symmetry-breaking potential (including the CW one) are otherwise inconsistent with observations.
Models like Ref. \cite{Iso:2014gka}, 
where the CW inflation with a Higgs linear term comes from the quark condensate, should be reanalyzed within the present framework.

(iv)  Another  consequence of the first-order QCD PT is that the formation of primordial black holes  (PBHs, see, e.g., Ref. \cite{Jedamzik:1996mr}) as well as 
of primordial magnetic fields~\cite{Sigl:1996dm} is eased.
If PBHs form, due to the horizon size at the QCD scale their mass is predicted in the (tenths of) solar mass range:
They might contribute to the dark matter of the Universe, can be searched for via lensing, and, being massive enough, through accretion disks they may alter the heating and ionization history between CMB recombination and first star formation, with consequences for CMB observables as well as for future 21 cm probes~\cite{Poulin:2017bwe}.

(v) The most direct cosmological probe would consist in the detection of the GW background produced via bubble collisions. 
Following the standard formulas \cite{Huber:2008hg,Binetruy:2012ze},
the GW power spectrum is determined by $\beta/{\cal H}$,
where ${\cal H}$ is the Hubble parameter at the production of GWs and $1/\beta$  corresponds to the duration of the PT and the typical size of the bubbles at the collision. 
The parameter  $\beta/{\cal H}$ is hard to compute reliably, 
although it is expected to be larger than $\sim$100 under reasonable assumptions~\cite{Hogan:1984hx}.
An additional parameter is $\zeta\equiv T_{\rm rh}/T_i$, with $T_{\rm rh}$ the reheating temperature, which quantifies the duration of the reheating period where the scalar oscillates around the true minimum behaving like pressureless matter.
In Fig.~\ref{Fig:GW}, we illustrate the approximate GW  signal expected under the assumption $T_i=10\, {\rm GeV}$
with varying $\zeta$ and $\beta/{\cal H}$. 
It is worth stressing that, in scenario (I), $\beta/{\cal H}$ is essentially independent from $g$,
in contrast to $\beta/{\cal H} \propto g^{-2}$ in scenario (II) \cite{V0-available}.
For comparison, the sensitivities of three configurations foreseen
for the space mission LISA~\cite{Caprini:2015zlo} are also reported. 
\\

\section*{Conclusions}
Phenomenologically, we know only that the thermal history of the Universe is conventional below temperatures of a few MeV, sufficient to set up the initial conditions (e.g., populating active neutrino species) for primordial nucleosynthesis.
It is usually assumed that the knowledge of the SM allows one to backtrack the evolution of the Universe up to temperatures of few hundreds of GeV, and that the EWPT is a crossover, as predicted by the SM, although theories with an extended EW sector where a first-order EWPT occurs are not rare.
It is, however, almost universally accepted that the QCD PT is not first-order, {\it even in models of physics beyond the SM}, hence with very limited implications for the later Universe.
Here we offer a counterexample, where an extension of the SM motivated only by EW physics sector changes both the QCD and EW PT dynamics, with the possibility of a very peculiar history of the Universe:
A first-order QCD PT (with six massless quarks) triggers a first-order EWPT, eventually followed by a low-scale reheating of the Universe where hadrons (likely) deconfine again, before a final, ``conventional'' crossover QCD transition to the current vacuum.
To the best of our knowledge, this is the {\it only viable scenario
known where a first-order QCD PT can be obtained} without large lepton~\cite{Schwarz:2009ii} or baryon asymmetry~\cite{Boeckel:2009ej}.
We have only sketched some important particle physics and cosmological consequences of this scenario. The actual reach of forthcoming collider searches, the extension to more general models than the $B-L$ here used for illustration, as well as quantitative consequences for cosmological crucial problems such as dark matter or baryon asymmetry 
are all interesting aspects which we plan to return to in the near future.\\

\section*{Acknowledments}
We thank N. Yamada and R. Jinno for useful discussions and G. Servant for reading the manuscript.
The work is supported by the Toshiko Yuasa France-Japan Particle Physics Laboratory (TYL-FJPPL) initiative.
S.I. is supported in part by JSPS KAKENHI Grant No. JP16H06490 and No. JP23540329.

\section*{Note added}
After the completion of this article, we became aware of an ongoing, related study~\cite{SvH}
motivated in the context of Randall-Sundrum models.

%%%%%%%%%%%%%%%%%%%%%%%%%%%%%%%%%%%%%%%%%%%%%%%%%%%%
%%%%%%%%%%%%%%%%%%%%%%%%%%%%%%%%%%%%%%%%%%%%%%%%%%%%
%%%%%%%%%%%%%%%%%%%%%%%%%%%%%%%%%%%%%%%%%%%%%%%%%%%%
%%%%%%%%%%%%%%%%%%%%%%%%%%%%%%%%%%%%%%%%%%%%%%%%%%%%

\newpage
\section*{Supplemental Material}

\subsection*{Zero-$T$ one-loop effective potential}
The tree-level part of the potential for the (physical components of) the $B-L$ Higgs $\phi = \sqrt{2} |\Phi| = \chi \cos \theta$ and the SM Higgs $h= \sqrt{2} |H| = \chi \sin \theta$ is given by
\be
V_{\rm tree} &=& \frac{\lambda_{\phi ,0}}{4} \phi^4 + \frac{\lambda_{{\rm mix},0}}{4} \phi^2 h^2 + \frac{\lambda_{h ,0}}{4} h^4 \ . \label{tree} \\
&=& \frac{\chi^4}{4} \left( \lambda_{\phi ,0} {\sf c}^4 + \lambda_{{\rm mix} ,0} {\sf c}^2 {\sf s}^2 +\lambda_{h ,0} {\sf s}^4 \right) \notag
\ee
where ${\sf c} \equiv \cos \theta$ and ${\sf s} \equiv\sin \theta$.
Following the Gildener-Weinberg formulation \cite{Gildener:1976ih}, the renormalization scale $\mu$ is chosen so that $V_{\rm tree}$ has a ``valley'' (flat direction at tree-level) along $\theta = \theta_0$ such that
\be
V_{\rm tree} \propto  \lambda_{\phi , 0} {\sf c}^4_{0}+  \lambda_{{\rm mix},0} {\sf c}_0^2 {\sf s}^2_{0} + \lambda_{h,0} {\sf s}^4_{0} = 0 \label{flat}
\ee
is satisfied with $\lambda_{{\rm mix},0}^2 = 4 \lambda_{h,0} \lambda_{\phi , 0}$.
Then we have
\be
\tan^2 \theta_0 = \frac{-\lambda_{{\rm mix} , 0}}{2 \lambda_{h,0}} \ . \notag
\ee
Let $\widetilde{m}_{i}(\chi , \theta)$ denote the mass of $i$ particle.
Provided that the scalar mixing coupling is small, $\lambda_{{\rm mix},0} \ll \lambda_{h , 0}$,
the canonically-normalized fluctuation in the $\theta $ direction can be identified as the almost SM Higgs boson $h$, with mass squared $\widetilde{m}_{h}^2(\chi,\theta_0) = |\lambda_{{\rm mix},0}| \chi^2$ in the valley.

In order to guarantee the stability of the  hierarchy between $M$ and $v$, 
the radiative corrections to the scalar mixing $\lambda_{\rm mix}$ should 
be much smaller than 1. The condition is given by
$ g_{\rm mix}^2 g^2 \ll 1 $ [21]. Thus it is satisfied as far as either
the gauge kinetic mixing $g_{\rm mix}$ or $g$ is small. 
Furthermore, since the radiative correction to
the gauge kinetic mixing is proportional to $g g_{\rm Y}^2$ where $g_{\rm Y}$ is
the hypercharge coupling, it is always subdominant to the $B-L$ coupling $g$ if we suppose that
the tree level mixing is small. 

The one-loop Coleman-Weinberg (CW) potential is given as a function of $\chi$ and $\theta$ by
\be
V_{\rm CW} &=& V_{\rm tree} +  \sum_{i}\delta_i V   + V_0\, , \notag
\ee
with
\be
\delta_i V(\widetilde{m}_i^2) = \sum_{i} \frac{(-1)^{2 s_i} n_i}{4\, (4 \pi)^2} \widetilde{m}_{i}^4  \ln \frac{\widetilde{m}_{i}^2}{\mu^2 e^{c_i}} \, , \notag
\ee
where $s_i$ and $n_i$ are the spin and the number of degrees of freedom of the particle $i$ and $c_i$ is the constant $3/2 \, (5/6)$ for $s_i = 0,\, 1/2$ ($s_i =1$) in $\overline{MS}$ scheme, and $\mu$ the renormalization scale.
The global minimum $(\chi , \theta) = (M_{\chi} , \theta_0 + \delta \theta)$ appears in the valley via the CW mechanism.
$V_0$ is introduced to cancel the potential energy there.
Imposing $\partial_{\chi } V_{\rm CW}=0$ at $(\chi , \theta) = (M_{\chi} , \theta_0 + \delta \theta)$, one obtains
\be
4 \sum_i \delta_i V /M_{\chi}^4 = - B /4 \ , \label{stationary}
\ee
\be
B = B_{\phi} {\sf c}_0^4 + B_{\rm mix} {\sf c}_0^2 {\sf s}_0^2 + B_h {\sf s}_0^4 \notag
\ee
where $B_{\phi}$, $B_{\rm mix}$ and $B_{h}$ are contributions from the vertex corrections to the beta-functions of the couplings $\lambda_\phi$, $\lambda_{\rm mix}$ and $\lambda_h$ respectively. 
For the small scalar mixing coupling, we have ${\sf c}_0^4 \approx 1+\lambda_{{\rm mix},0}/\lambda_{h,0}$ and
\be
B &\approx &\frac{{\sf c}_0^4}{8 \pi^2} \left(3 (2 g)^4 - 2{\rm Tr}\left(\frac{Y}{\sqrt{2}}\right)^4 - D \, \lambda_{\rm mix}^2 \right)
 \ ,  \notag
\ee
\be
D = - \frac{m_h^4 -12 m_t^4 + 3 m_{Z}^4 + 6 m_{W}^4 + \cdots}{m_h^4} \simeq 41 \ . \notag
\ee
The explicit value of $\delta \theta$ follows from $\partial_{\theta} V_{\rm CW}=0$ as $\delta \theta = - (\partial_{\theta} \sum_i \delta_i V ) /\partial_{\theta}^2 V_{\rm tree}|_{\chi = M_{\chi}, \theta = \theta_0}$.
And the vacuum expectation values of $\phi$ and $h$ are given by
\be
M &=& M_\chi \cos (\theta_0 + \delta \theta) \approx M_{\chi} \ , \notag \\
v &=& M_\chi \sin (\theta_0 + \delta \theta) \approx \sqrt{|\lambda_{\rm mix}|/(2 \lambda_{h})} \times M_{\chi} \ . \notag
\ee
at the stationary point.
With the condition (\ref{stationary}), we obtain the simple form of the scalar potential along the valley as
\be
V_{\rm CW}^{\rm v} = V_0 + \frac{B}{4} \chi^4 \left( \ln \frac{\chi}{M} -\frac{1}{4} \right) \ . \notag
\ee
The constant $V_0$ is fixed as $V_0 = B M^4 /16$.
The stability condition $B>0$ for the CW mechanism to work is
\be
m_{Z'}^4 >\frac{2 {\rm Tr}\, m_{N}^4}{3} + \frac{D\, m_h^4}{3} \ . \label{B>0}
\ee

%%%%%%%%%%%%%%%%%%%%%%%%%%%%%%%%%%%%%%%%%%%%%%%%%%%
\subsection*{Finite-$T$ effective potential}
The resumed one-loop correction to Eq.~(\ref{tree}) at finite-$T$ is formally given by
\be
\Delta V_{T} &=& \sum_{i} \left\{ \delta V_i(\widetilde{m}_{{\rm eff},i}^2)+ (-1)^{2 s_i} n_i \frac{T^4}{2 \pi^2} I_i(\widetilde{m}_{{\rm eff},i}^2) \right\} \ , \notag
\ee
\be
I_i(\widetilde{m}_{{\rm eff},i}^2) = \int dy \, y^2 \ln \left( 1- (-1)^{2 s_i} e^{-\sqrt{y^2 + \widetilde{m}^2_{{\rm eff},i}/T^2}} \right) \ .\notag
\ee
Notice that the effective mass squared $\widetilde{m}_{{\rm eff},i}^2$ includes the thermal contribution $\propto T^2$ in a self-consistent way \cite{Parwani:1991gq}.

In order to make the condition (\ref{B>0}) satisfied, we assume $m_{Z'} \gtrsim 1.9\,m_h$,
then in the $\phi$ direction the $Z'$ loop correction dominates the thermal potential.
For simplicity we assume $m_{N_j}\ll m_{Z'}$, neglecting the RH$\nu$ contribution. 
On the other hand, the thermal potential in the $h$ direction is dominated by the SM particles.
The effective Higgs mass is given by
\be
\widetilde{m}_{{\rm eff},h}^2 = 3 \lambda_h h^2 -\frac{|\lambda_{\rm mix}|}{2}\phi^2 + A T^2\,,  \label{m_h,eff}
\ee
with $A=\{m_t^2/2 + (2 m_W^2 +m_Z^2)/4 +\cdots \} /v^2 \sim 1/4$.
Eq.~(\ref{m_h,eff}) shows that, for $\phi^2 < \phi_{\rm EWSB}^2 \equiv T^2 2 A /|\lambda_{\rm mix}|$,
{\it the potential valley lies along} $h=0$, where the SM Higgs boson has effective mass $\widetilde{m}_{{\rm eff},h}^2 = A T^2 -|\lambda_{\rm mix}| \phi^2 /2$ and the thermal potential is given by
\be
V_{T}^{\rm v} &=& \frac{\lambda_{\phi ,0}}{4} \phi^4  + \Delta V_T |_{h=0} \notag
\ee
with $\lambda_{\phi , 0} = \lambda_{{\rm mix},0}^2 / (4 \lambda_{h , 0})= \lambda_{h , 0} \tan^2 \theta_0$, see below (\ref{flat}).

\begin{figure}[t]
\centering
%\vspace{-4mm}
\includegraphics[width=1 \linewidth, bb=0 0 980 394]{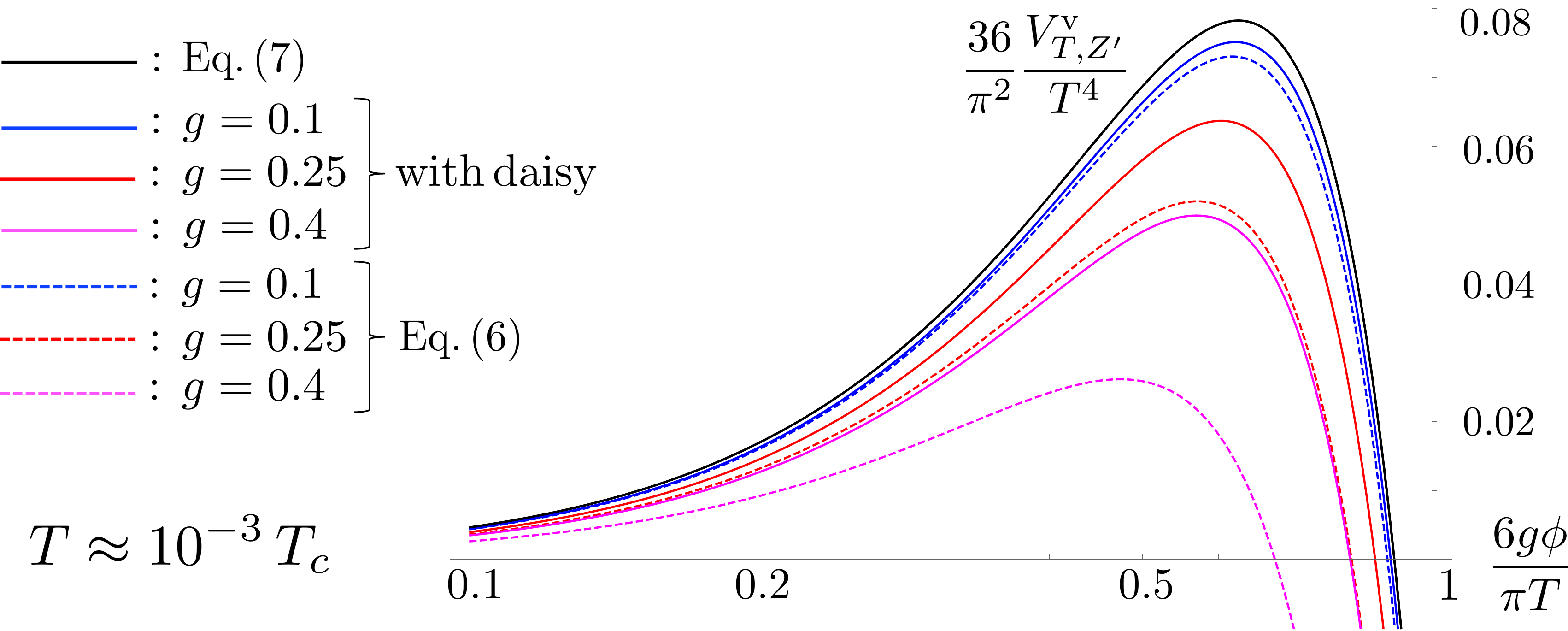}
\caption{\label{Fig:suppl}
Comparison of Eq.(\ref{leading}), Eq.(\ref{leading})$+V_{T,Z'}^{\rm daisy}$ and Eq.(\ref{thermal-potential}).}
\end{figure}

In the simplified analysis in the section {\it Hypercooling in the EW sector}, we take $\theta_0 \to 0$ limit and consider only the $Z'$ contribution:
\be
V_{T}^{\rm v} \to V_{T,Z'}^{\rm v} \equiv  \sum_{k={\rm T, L}} \left\{ \delta V_{Z',k} + n_k \frac{T^4}{2 \pi^2} I_{Z',k}  \right\} \label{thermal-potential}
\ee
where
the effective mass of the $Z'$ boson is given by
\be
\widetilde{m}_{{\rm eff},Z',{\rm T}}^2 = 4 g^2 \phi^2 \ , \ \ \widetilde{m}_{{\rm eff},Z',{\rm L}}^2 = 4 g^2 \phi^2 + 4 \, g^2 \, T^2 \notag
\ee
for the two transverse modes and the longitudinal mode, respectively (see \cite{Arnold:1992rz} for a chiral Abelian Higgs model).
Note that a condition $ (3 m_{Z'}^4 /(4 \pi)^2 M^4) \ln(m_{Z'}^2 /\mu^2 e^{5/6}) = - B_{Z'}/4$ corresponding to (\ref{stationary}) is imposed here with the definition $B_{Z'} \equiv 3 (2 g)^4 /8 \pi^2$.
Furthermore, we work with the high-$T$ approximation ($\widetilde{m}_{{\rm eff},Z'} < T$) neglecting the so-called daisy corrections, $V_{T,Z'}^{\rm daisy} = -(c_3/9) T \{ \left( \phi^2 +T^2 \right)^{3/2} - \phi^3 \}$:
\begin{equation}
V_{T,Z'}^{\rm v} \approx \frac{c_2}{2}T^2 \phi^2 -\frac{c_3}{3} T \phi^3 + \frac{B_{Z'}}{4} \phi^4 \ln \frac{T}{\hat{\mu}}\,, \label{leading}
\end{equation}
with the parameters $c_2= g^2$, $c_3 = 6 g^3 /\pi$,
$\hat{\mu}= m_{Z'} e^{\gamma_{E}-1/2}/(4 \pi)$. (We omit terms which are $\phi$-independent or higher order in couplings.)
At $T= e^{3/2}\hat{\mu}$, the non-trivial minimum appears, and eventually at $T_c = e^{4/3} \hat{\mu}$ it acquires lower energy compared to the one at the origin.
At $T<T_c$, the thickness of the potential barrier defined by Eq.~(\ref{leading})=0  is given by $\Delta \phi \approx (T/g) \times 2 \sqrt{3/\ln (T/\hat{\mu})}$, which is smaller than $\phi_{\rm EWSB}$ and the analysis in the valley along $h=0$ is sufficient.
The height of the barrier is roughly given by $c_2 T^2 \Delta \phi^2 \sim T^4/\ln (T/\hat{\mu})$. We checked numerically that, for small coupling $g\lesssim 0.1$, Eq.~(\ref{leading}) provides a decent approximation to  the full thermal potential of Eq.~(\ref{thermal-potential}) well below $T_c$, while adding $V_{T,Z'}^{\rm daisy}$ only marginally changes the curve. This approximation is however poor when the coupling becomes large, even if the daisy correction are included. The situation for the two cases is summarized in Fig.~\ref{Fig:suppl}. We conclude that the approximation used in the section {\it Hypercooling in the EW sector} is qualitatively correct, and  hypercooling takes place if the QCD dynamics is omitted. Yet, quantitatively the  contour lines of $T_p$ in Fig. 1 should move toward smaller $g$ when the daisy correction or the full thermal potential (\ref{thermal-potential}) is considered.
%%%%%%

%\vspace{-3mm}
If we keep $\theta_0$ finite and consider the case of $m_{Z'} \sim m_{h}$ where the stability condition of Eq.~(\ref{B>0}) is only barely fulfilled (note that we don't assume strong cancelation in $B$),
$\Delta \phi$ is as large as $\phi_{\rm EWSB}$ beyond which the SM Higgs direction becomes tachyonic along $h=0$.
Although the situation in the $\phi-h$ plane is now modified, we expect that our conclusions are qualitatively unchanged. The width of the barrier is at least as large as $\phi_{\rm EWSB}\sim T/\sqrt{|\lambda_{\rm mix}|} \sim T/g$, while the height of the barrier, estimated by replacing this value in the potential, is still of order $T^4$.
Hence the tunneling rate is comparable or lower than the one estimated in the main text and even near the lower bound on $m_{Z'}$ given by Eq.~(\ref{B>0}), the nucleation/percolation temperature is as low as what computed in our simplified analysis.


\begin{thebibliography}{99}
\bibitem{V0-available}
If the slow roll parameter $m_{\rm pl}^2 |V''| / V_0$ is smaller than $(\beta/{\cal H})^2$, 
the scalar field, starting from the false vacuum, cannot reach at 
the true minimum before bubbles collide.
Thus the vacuum energy cannot be fully released to the GWs, and their amplitude is reduced.
\end{thebibliography}
\end{document}